\documentclass[11pt]{article}

\pdfoutput=1

\usepackage{amssymb,amsfonts,fullpage,graphicx,nicefrac,slashed,tikz,latexsym}
\usepackage{graphicx}
\usepackage{epsfig}
\usepackage{color} 
\usepackage{braket}
\usepackage[english]{babel}
\usepackage[compress]{cite}
\usepackage[intlimits]{amsmath}
\usepackage[normalem]{ulem} 
\usepackage[normal,font=small,labelfont=bf,labelsep=period]{caption}
\usepackage{fullpage}
\usepackage{color}
\definecolor{colorLink}{rgb}{0.9,0,0} 
\definecolor{colorCite}{rgb}{0,0,0.7} 
\definecolor{colorURL} {rgb}{0,0,0.8} 
\definecolor{colorCapt}{rgb}{0.5,0,1} 
\definecolor{colorPink}{rgb}{1,0,0.7} 
\usepackage[colorlinks=true,linktocpage=true,linkcolor=colorLink,citecolor=colorCite,urlcolor=colorURL]{hyperref}

\title{\bf Stationary state after a quench\\ to the Lieb-Liniger from rotating BECs}
\author{Leda Bucciantini$^{1}$}
\date{\it \small Dipartimento di Fisica dell'Universit\`a di Pisa and INFN, 56127 Pisa, Italy\\
\small Max Planck Institute for the Physics of Complex Systems, 01187 Dresden, Germany}

\begin{document}
\begin{titlepage}
\maketitle
\thispagestyle{empty}

\begin{abstract}
\noindent
We study long-time dynamics of a bosonic system after suddenly switching on repulsive delta-like interactions. As initial states, we consider two experimentally relevant configurations:  a rotating BEC and two counter-propagating BECs with opposite momentum, both on a ring. In the first case, the rapidity distribution function for the stationary state is derived analytically and it is given by the distribution obtained for the same quench starting from a BEC, shifted by the momentum of each boson. In the second case, the rapidity distribution function is obtained numerically  for generic values of repulsive interaction and initial momentum. The significant differences for the case of large versus small quenches are discussed.

\end{abstract}

\vfill
\noindent\line(1,0){188}\\\medskip
\footnotesize{E-mail: \scriptsize\tt{\href{mailto:lbucci@pks.mpg.de}{lbucci@pks.mpg.de}}}

\end{titlepage}



\section{Introduction}\label{intro}
In the last ten years, out-of-equilibrium many-body quantum physics has become a very active area of research, especially fuelled by the experimental realizations developed in  cold-atomic setups \cite{RevModPhys.80.885,RevModPhys.83.1405,Greiner2002,PhysRevLett.81.3108}. As opposed to usual solid state devices, with ultracold atoms it is  possible to experimentally
realize quantum systems in which the effect of decoherence and dissipation is negligible on long
time scales \cite{Kinoshita,Sadler2006,Trotzky2012,Gring1318,Schneider2012,Hofferberth2007}. Essentially unitary time evolution of an initially isolated system can be studied,
enough for allowing the observation of the stationary configuration eventually reached \cite{PhysRevLett.100.100601, PhysRevLett.100.030602,1367-2630-12-5-055020}, if any. This has led to the possibility of
addressing long standing theoretical questions concerning equilibration and thermalisation in many body quantum systems, 
the role of integrability and universality  and the study of the time-dependence of correlation functions \cite{RevModPhys.83.863,doi:10.1080/00018732.2010.514702,2010arXiv1012.0653D,Rigol2008,Bloch2008}.

One of the most studied out-of-equilibrium protocol is the sudden quantum quench \cite{PhysRevLett.96.136801,1742-5468-2007-06-P06008,PhysRevA.69.053616,PhysRevLett.93.142002,PhysRevLett.97.156403,PhysRevB.74.174508,PhysRevLett.95.105701,PhysRevB.79.155104}. After having prepared a system in an initial configuration, eigenstate of a pre-quenched Hamiltonian, from a certain instant of time on it is let evolved isolated from the rest according to a different Hamiltonian (the post-quenched one). In this kind of settings, the main focus is the investigation of the properties of the stationary state reached long after the quench. To make the time evolution non trivial, the initial state should not be chosen as an eigenstate of the post-quenched Hamiltonian. When considering a thermodynamically large system, this initial state has a non vanishing overlap with a very large number of eigenstates of the post-quenched Hamiltonian. If the thermodynamic limit is taken before the long time limit, a dephazing mechanism takes place: as a consequence, any observable relative to a finite subsystem  will relax to time-independent values, even if the entire system is isolated \cite{1367-2630-12-5-055020}. 

While for non integrable systems the stationary distribution is generally given by  the thermal Gibbs one \cite{PhysRevA.43.2046,PhysRevE.50.888}, it has been conjectured that integrable systems relax to a Generalized Gibbs Ensemble (GGE) \cite{PhysRevLett.98.050405}, obtained imposing  the conservation of all local  charges  throughout the unitary evolution. For free, or mappable-to-free systems, the conserved charges can be identified with the momentum occupation numbers and thus the GGE can be explicitly constructed \cite{PhysRevLett.106.227203,2012JSMTE..07..016C,2012JSMTE..07..022C,PhysRevB.84.212404,PhysRevLett.102.127204,2011PhRvL.106c5701I,1742-5468-2012-04-P04017,PhysRevLett.107.150602,1742-5468-2014-7-P07024}, \cite{2014JPhA...47q5002B,2014EL....10740002K}. In the case of a truly interacting (yet integrable) model, i.e. for those having a dressed two-body scattering matrix, the construction of the GGE is more involved, due to the difficulty in writing the conserved charges in a closed operatorial form \cite{2010NJPh...12e5015F,PhysRevLett.109.175301,2013JSMTE..07..012F,PhysRevB.89.125101,2013JSMTE..07..003P}. As pointed out in \cite{PhysRevB.88.205131}, the GGE is hard to construct for the Lieb Liniger model because the conserved charges are divergent. Moreover, it has been recently pointed out that a GGE which takes  into account only the local conserved charges is not able to capture the stationary state in some cases \cite{PhysRevLett.113.117202,PhysRevLett.113.117203}, \cite{2015JSMTE..04..001M},\cite{2014JSMTE..09..026P} and  one needs to include also the quasi local charges \cite{PhysRevA.91.051602,PhysRevLett.115.157201,2015arXiv150706994A}.

It has  been shown that there is another way of computing expectation values of local quantities at any time after a quench in  integrable models, the so-called quench action method \cite{PhysRevLett.110.257203}. In particular, it is proved that the long time behaviour of local observables is given by the expectation value with respect to a single representative state of the (post-quenched) Hamiltonian. This representative state can be obtained as the result of a variational method, exact in the thermodynamic limit, which requires the knowledge  of the overlap of any eigenstate of the post-quenched Hamiltonian with the initial state. As a consequence, this method is potentially very useful for interacting integrable models \cite{2015JSMTE..04..001M, PhysRevA.89.033601, 2014JSMTE..12..009B,2016PhRvL.116g0408P,PhysRevA.91.021603,2014JSMTE..10..035B} because it does not require having the expression of the conserved charges, differently from the GGE approach. Nonetheless, the expression for the overlaps is often highly non trivial \cite{2014JPhA...47n5003B,2014JPhA...47H5003B} and always initial state dependent.

In the present paper, by making use of the quench action method, we solve two quench problems that have recently attracted the attention of  both experimentalists and theoreticians \cite{Kinoshita,2015PhRvL.115q5301V,PhysRevLett.92.050403,RevModPhys.81.647,doi:10.1080/00018730802564122,PhysRevA.88.063633,PhysRevLett.102.155301,PhysRevLett.109.115304,1742-5468-2010-05-P05012,2015PhRvB..92o5103G,2015arXiv150706339V}.  We obtain results for the steady-state reached by an initially free bosonic system after switching on at $t=0$ repulsive  contact interaction  (Lieb-Liniger model \cite{2012NJPh...14g5006M,1751-8121-48-43-43FT01,2014JSMTE..12..012D,PhysRevLett.111.100401,2014JSMTE..01..009C}). As initial states, we consider two experimentally relevant configurations, that are both eigenstates of the Lieb-Liniger model at zero interaction strength. The first is a rotating BEC and the second consists of many couples of counter propagating BECs with opposite momentum, moving on a ring. In this latter case, our calculations represent a close theoretical description of the cornerstone Quantum Newton's Cradle \cite{Kinoshita}. The only differences are that we do not take into account the presence of the trap \cite{2014arXiv1407.1037M} and that we actually perform a quench of the interaction strength $c$, from $c=0$ at $t=0$ to $c>0$  at $t>0$, while in the actual experiment the two counter-propagating wavepackets are prepared and let evolved at a fixed value of $c$. \\
We note that these also represent physical examples of a quench where the initial state is an excited state of the pre-quenched Hamiltonian \cite{2014JPhA...47q5002B,2014EL....10740002K}, depending on the value of the momentum. This is   different from the majority of quenches studied in literature, where the initial state is the ground state of the pre-quenched Hamiltonian. 

In the case of the initially rotating BEC, we have found that the global shape of the rapidity root distribution in the asymptotic state is not affected by the rotation of the BEC and it is just shifted uniformly with respect to the static BEC case by the momentum of each boson. For the initially counter propagating BECs, instead, the  entire structure of the root distribution is modified and it is very different in the large and small quench limit.
  
The paper is organized as follows: in Sec.  \ref{model} and \ref{quenchactionmethod}   we  review the main features of the Lieb-Liniger model and the quench action method while in the remaining sections we present our original results. In Sec. \ref{rotBEC} we present the results for the rapidity distribution of the asymptotic state after a quench  from  the initial rotating BEC. In Sec. \ref{KINO} we obtain the stationary root distribution after a quench from the colliding BECs and  we discuss the approximate solutions in the case of a large and small quench. In Sec. \ref{concl} we draw conclusions.


\section{The model}\label{model}
In this section we review the basics of the Lieb-Liniger model \cite{PhysRev.130.1605,PhysRev.130.1616} and set our conventions, following \cite{PhysRevA.89.033601}. We consider a single component bosonic quantum gas of $N$ particles in a 1D box of length $L$ with periodic boundary conditions.  From $t=0$ the system evolves unitarily  according to the Lieb-Liniger Hamiltonian \cite{PhysRev.130.1605,PhysRev.130.1616} (setting $\hslash=2 m=1$)
\begin{equation}\label{HLL1}
H_{LL}(c)=-\sum_{i=1}^N\frac{\partial ^2}{\partial x_i^2}+2c \sum_{i<j}^N\delta(x_i-x_j).
\end{equation}
The parameter $c$ describes the strength of interaction and we will consider $c>0$ in the rest of the paper, so as to depict repulsion between particles. This is a realistic model for interacting bosons in quantum degenerate gases with $s$-wave scattering potential. 

The model can be solved by means of the Bethe Ansatz \cite{Bethe1931}. The solution of (\ref{HLL1}) can be found taking into account   the symmetry of the bosonic wavefunction under two particles' exchange and  the fact that the $N$ particles are free unless two of them occupy the same position. Accordingly, we can consider all the possible $N!$ domains defined by the ordered non-coinciding particles' positions $\Theta({\cal P}): 0\leq x_{{\cal P}_1}< x_{{\cal P}_2}< \dots <x_{{\cal P}_N }\leq L$, where ${\cal P}$ is the permutation of the number set $\{1,2, \dots, N\}$. The total wavefunction can be written as the sum of the wavefunctions in each domain
\begin{equation}
\psi(\boldsymbol{x})=\sum_{\cal P} \Theta({\cal P}) \psi_{\cal P}(\boldsymbol{x}),
\end{equation}
where $\psi(\boldsymbol{x})$ is the bosonic wavefunction written in first quantization. Due to the Bose statistics the wavefunction is the same in all  domains: $\psi_{\cal P}=\psi_{\boldsymbol{1}}$, where $\boldsymbol{1}=\{1,2,\dots ,N\}$ indicates the domain $\{0\leq x_1<x_2<\dots <x_N\leq L\}$. 
For the model ($\ref{HLL1}$), $\psi_{\boldsymbol{1}}$ can be written as a superposition of $N!$ plane waves 
\begin{equation}
\psi_{\boldsymbol{1}}(\boldsymbol{x})=\sum_{P\in S_N} S (P)   \prod_{j=1}^N e^{i \lambda_{P_j} x_j},
\end{equation}
where $S_N$ are all the possible permutations of the set of rapidities  $\boldsymbol{\lambda}\equiv\{\lambda_1, \lambda_2,\dots, \lambda_N \}$ among the particles at position $\boldsymbol{x}=\{x_1, x_2, \dots, x_N\}$. The coefficients $S(P)$ can be found imposing the order of the particles in domain ${\boldsymbol{1}}$ in   the Schro\"edinger equation, i.e. $(\partial_{x_{j+1}}-\partial_{x_j}-c)\psi(\boldsymbol{x})|_{x \in \partial _{\boldsymbol{1}}}=0$, where $\partial _{\boldsymbol{1}}: x_{j+1}=x_{j} + 0^+$, $\forall j=1, \dots N-1$.  A generic (not normalized) eigenstate of (\ref{HLL1}) with a given set of rapidities $\boldsymbol{\lambda}$ has the form \cite{PhysRevA.89.033601}
\begin{equation}\label{HLL}
\psi(\boldsymbol{x}|\boldsymbol{\lambda})=\langle\boldsymbol{x}|\boldsymbol{\lambda}\rangle =F_{\boldsymbol{\lambda}} \sum_{P\in S_N} A_P(\boldsymbol{x}|\boldsymbol{\lambda}) \prod_{j=1}^N e^{i \lambda_{P_j} x_j},
\end{equation}
with
\begin{equation}
F_{\boldsymbol{\lambda}}=\frac{\prod_{j>k=1}^N (\lambda_j -\lambda_k)}{\sqrt{N! \prod_{j>k=1}^N(c^2+(\lambda_j -\lambda_k)^2)}}, \qquad A_P(\boldsymbol{x}|\boldsymbol{\lambda})=\prod_{j>k=1}^N \left(1-\frac{ic~ \mathrm{sgn}(x_j-x_k)}{(\lambda_{P_j}-\lambda_{P_k})}\right).
\end{equation}
The norm of the wavefunction is given by the Gaudin determinant \cite{1993qism.book.....K}.

Imposing periodic boundary conditions, the rapidities get quantized and have to satisfy the Bethe equations \cite{Bethe1931}, a set of $N$ coupled algebraic equations
\begin{equation}\label{BE_discre}
\lambda_j=\frac{2 \pi I_j}{L}-\frac{2}{L}\sum_{k=1}^N \mathrm{arctan}\left(\frac{\lambda_j-\lambda_k}{c}\right) \qquad j=1\dots N, 
\end{equation} 
where $\boldsymbol{I}\equiv \{I_1, I_2, \dots, I_N\}$ are the quantum numbers of the rapidities, which label an eigenstate uniquely and are $c$-independent, differently from $\boldsymbol{\lambda}$. The full set of allowed quantum numbers is the union of the occupied ones $\boldsymbol{I}$ and the unoccupied ones, defined as $\boldsymbol{I_h}$. If $N$ is odd, $\boldsymbol{I}$ are the integers and $\boldsymbol{I_h}$ the semi-integers; if $N$ is even the vice-versa. 

The solution of the Bethe equations (\ref{BE_discre}) provides an exhaustive knowledge of the spectrum of the Lieb-Liniger model; for a given set of rapidities $\boldsymbol{\lambda}$, the total momenta and energy of the system are obtained as
\begin{equation}\label{E_P_discre}
P=\sum_{j=1}^N \lambda_j, \quad E=\sum_{j=1}^N \lambda _j^2 .
\end{equation}
Higher conserved charges $Q_n$, $n \in \mathbb{N}$, are of the form\cite{DAVIES1990433,2011arXiv1109.6604D}
\begin{equation}
Q_n=\sum_{j=1}^N \lambda _j^n .
\end{equation}
As already mentioned in Sec. \ref{intro}, in our quench protocol we are interested in the thermodynamic limit of this model ($N \to \infty$, $L \to \infty$, with $N/L$ fixed). It is useful to introduce the (particle) root density $\rho(\lambda)$, i.e. the  density distribution function of the rapidities defined by the particle numbers $\boldsymbol{I}$ in the interval  ($\lambda$, $\lambda+\Delta \lambda$)
\begin{equation}\label{defroot}
\rho(\lambda)\equiv \lim_{\Delta \lambda\to 0} \frac{1}{L \Delta \lambda}.
\end{equation}
Accordingly we can define a hole root density $\rho_h(\lambda)$, i.e. a density distribution function of the rapidities defined by $\boldsymbol{I_h}$, which are the unoccupied quantum numbers of the allowed set, in the interval  ($\lambda$, $\lambda+\Delta \lambda$).  
Hence the Bethe equations (\ref{BE_discre})  can be rewritten in the thermodynamic limit as
\begin{equation}\label{BE_th}
\rho(\lambda)+\rho_h(\lambda)=\frac{1}{2 \pi} + \int_{-\infty}^{\infty} \frac{d \mu}{2 \pi} K(\lambda-\mu)\rho(\mu),
\end{equation}
where 
\begin{equation}\label{K_fun}
K(\lambda)=\frac{2c}{\lambda^2+c^2}.
\end{equation}
The momentum and energy of the system per unit length (\ref{E_P_discre}) thus become 
\begin{equation}
\frac{P}{L}=\int_{-\infty}^{\infty}d\lambda \, \lambda \, \rho(\lambda), \quad \frac{E}{L}=\int_{-\infty}^{\infty}d\lambda\, \lambda^2 \rho(\lambda).
\end{equation}


\section{Quench action method: overview}\label{quenchactionmethod}
In this section, following \cite{PhysRevLett.110.257203}, we give an overview of the quench action method that we will use to obtain the results for  the steady states. This method provides a way of computing the expectation value  of a local observable at any time during the unitary evolution following a quench, in  a thermodynamically large system. We will show that the long-time behaviour   is given by the expectation value on a single eigenstate of the post-quenched Hamiltonian. This representative state can be simply computed extremizing a particular functional of the root density $\rho(\lambda)$, called quench action.

The aim is to compute the expectation value of a generic local observable $O$ at time $t$
\begin{equation}\label{a1}
\langle O(t)\rangle \equiv \frac{ \langle \psi(t)|O|\psi(t)\rangle}{\langle \psi(t)|\psi(t)\rangle},
\end{equation}
with $|\psi(t)\rangle$ being the time evolved state with the post-quenched Hamiltonian $H$ from the initial state $|\psi(0)\rangle$
\begin{equation}\label{unit_evo}
 |\psi(t)\rangle =e^{-iH t} |\psi(0)\rangle.
\end{equation}

Let us first focus on the numerator of (\ref{a1}); expanding it on the post-quenched eigenbasis $|I\rangle$ we get
\begin{equation}\label{a2}
\langle \psi(t)|O|\psi(t)\rangle= \sum_{I, I'}e^{-S_I^*-S_{I'}} e^{i(\Omega_I-\Omega_{I'})t} \langle I|O|I'\rangle,
\end{equation}
where $ S_I\equiv -\ln \langle I|\psi(0)\rangle$ and $H|I\rangle=\Omega_I |I\rangle$ .
The practical difficulty in evaluating such a double sum can be overcome by going to the thermodynamic limit. Any sum over a discrete set of quantum numbers can be transformed into a functional integral over all the microscopic configurations sharing the same root density $\rho(\lambda)$, each of which weighted with the Yang-Yang entropy $S_{YY}[\rho, \rho_h]$ \cite{1664947}
\begin{equation}\label{conti}
\lim_{\mathrm{th}}\sum_{I'}(\cdots) \to \int D[\rho, \rho_h] e^{S_{YY}[\rho, \rho_h]} (\cdots).
\end{equation}
The Yang-Yang entropy $S_{YY}[\rho, \rho_h]$  is defined  \cite{1664947} as 
\begin{equation}\label{YYeq}
S_{YY}[\rho, \rho_h] \equiv L\int_{-\infty}^{+\infty}d \lambda ([\rho(\lambda)+\rho_h(\lambda)]\ln[\rho(\lambda)+\rho_h(\lambda)]-\rho(\lambda) \ln \rho(\lambda)-\rho_h(\lambda)\ln \rho_h(\lambda)),
\end{equation}
where the hole density $\rho_h(\lambda)$ is related  to the particle density $\rho(\lambda)$ by the Bethe equations (\ref{BE_th}). To extract the leading behaviour of (\ref{a2}) in the thermodynamic limit, it is sufficient to transform only one of the two sums  in its continuum form with eq. (\ref{conti}). In fact, since $O$ is local, $\langle I|O|I'\rangle$ is non vanishing only provided that the states $|I\rangle, |I'\rangle $ have the same (macroscopic) root density in the thermodynamic limit.  After enforcing (\ref{conti}), (\ref{a2}) becomes
\begin{equation}\label{a3}
\langle \psi(t)|O|\psi(t)\rangle =\int {\cal D} [\rho] e^{S_{YY}[\rho]} \sum _I \left[e^{-i t(\Omega_\rho -\Omega _I)} \frac{\langle I|O|\rho \rangle}{2} e^{-S^*_I -S_{\rho}}+ e^{-i t(-\Omega_\rho +\Omega _I)} \frac{\langle \rho | O|I \rangle}{2} e^{-S_I -S_{\rho}^{*}}\right],
\end{equation}
where the integration and functional dependence on $\rho_h$ is kept implicit from now on. The states $|I\rangle$ and $|\rho\rangle$ can  only differ for a small number of microscopic particle-hole excitations. This yields $\sum_I \langle  I|=\sum_e \langle \rho +e|$, $\Omega_I\simeq \Omega_\rho+ \delta \omega_e$ and $S_I^*+S_{\rho}\simeq 2 \operatorname{Re}(S_\rho)+ \delta s_e^*$. Notice that $\delta\omega_ e$ and $\delta s_e$ are the microscopic differences in energy and overlap between $|\rho+e\rangle$ and $|\rho\rangle$, while $S_\rho$ and $\Omega_\rho$ are respectively  the extensive part of the overlap and energy,  in the thermodynamic limit. Eq. (\ref{a3}) can thus be rewritten as 
\begin{equation}\label{a4}
\langle \psi(t)|O|\psi(t)\rangle =\int {\cal D} [\rho] e^{S_{YY}[\rho]-2 \operatorname{Re}(S_\rho)} \sum _e \left[e^{i t  \delta \omega_e - \delta s_e^*} \frac{\langle \rho +e|O|\rho \rangle}{2} + e^{-i t  \delta \omega_e - \delta s_e} \frac{\langle \rho | O| \rho +e\rangle}{2}\right].
\end{equation}
This integral can be computed with a saddle point evaluation in the thermodynamic limit, since $S_{YY}[\rho]$  and $\operatorname{Re}(S_\rho)$ are  extensive in the system's size. The most significant contribution is given by the saddle point  distribution $\rho_s$ that makes the exponent in (\ref{a4})  stationary. Defining the quench action as
\begin{equation}\label{QAeq}
S^{QA}[\rho]\equiv 2 \operatorname{Re}(S_\rho)-S_{YY}[\rho],
\end{equation} 
$\rho_s$ is identified by 
\begin{equation}\label{sp_condi}
\left.\frac{\delta S^{QA}[\rho]}{\delta \rho}\right|_{\rho_s}=0.
\end{equation}


Let us now focus on the denominator of (\ref{a1}); its leading contribution is  given by the same root density $\rho_s$ of Eq. (\ref{sp_condi})
\begin{equation}
\langle\psi(t)|\psi(t)\rangle=\sum_I e^{-2 \operatorname{Re}(S_I)}=\int  D[\rho] e^{S_{YY}[\rho]-2 \operatorname{Re}[S_\rho]}.
\end{equation}
Putting everything together, the expectation value  (\ref{a1}) becomes
\begin{equation}\label{Ot}
\lim_{N\to \infty}\langle O(t)\rangle =\sum _e \left[e^{i t  \delta \omega_e - \delta s_e^*} \frac{\langle \rho_s +e|O|\rho_s \rangle}{2} + e^{-i t  \delta \omega_e - \delta s_e} \frac{\langle \rho_s | O| \rho_s +e\rangle}{2}\right].
\end{equation}
Eq. (\ref{Ot}) means that in the thermodynamic limit $\langle O(t)\rangle$ is fully determined by states whose root density distribution differs only by few microscopic  particle-hole excitations from the saddle point one.
 
When $t \to \infty$ a stationary phase approximation yields the additional condition
\begin{equation}
\frac{\delta(\Omega_ I -\Omega_{\rho})}{\delta \rho}=0,
\end{equation}
so that (\ref{Ot}) becomes
\begin{equation}
\lim_{t \to \infty}\lim_{N \to \infty}\langle O(t)\rangle = \langle \rho_s|O|\rho_s\rangle.
\end{equation}
The conclusion is that only one state matters for the stationary behaviour: it is the one fixed by the stationarity of the quench action (\ref{QAeq}).


\section{Quench from a rotating BEC to the Lieb-Liniger}\label{rotBEC}
In this section we present the first of our original contributions: the solution of the stationary state reached after a quench from $H_{LL}(c=0)$ to $H_{LL}(c>0)$ starting from  a rotating BEC. 

\begin{figure}[t] 
\begin{center} 
\includegraphics[width=.3\textwidth]{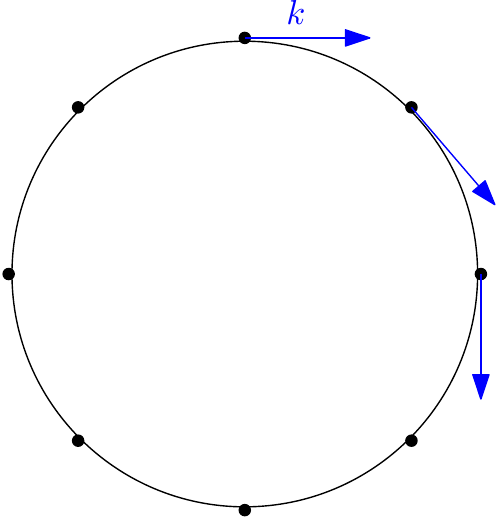}
\caption{\label{fig:figuraillustr_rotBEC}\small Schematic description of the rotating BEC. All bosons have the same value of momentum $k$ and lie on a ring.}
\end{center} 
\end{figure}

We represent the rotating BEC as a system of $N$ bosons on a ring of circumference $L$,  each of which with a given value of the momentum $k$, depicted in  Fig. \ref{fig:figuraillustr_rotBEC}.

 The state is given by
\begin{equation}\label{IS}
\langle {\bf x}|0^\rightarrow\rangle=\frac{1}{L^{N/2}}\prod_{i=1}^N e^{i k x_i},
\end{equation}
where $x_i$ is the position of the $i$-th boson. The rotating BEC is an excited eigenstate of $H_{LL}(c=0)$. At $t=0$ we suddenly turn on the interaction  to an arbitrary positive value and then let the system evolve unitarily until a steady state is reached.

The first step for obtaining the saddle point root distribution in the stationary state  is to construct  the quench action (\ref{QAeq}). It entails the   computation of the overlap between a generic Bethe state, eigenfunction of the post-quenched Lieb-Liniger Hamiltonian, and the initial state (see Appendix \ref{AppA} for the detailed computation). Due to momentum conservation, a non-vanishing overlap arises only for the Bethe states with rapidities that are symmetrically distributed around $k$. The saddle point evaluation (\ref{a4}), on which the quench action is based, implies that only the extensive  part of the overlap $S_\rho$ is relevant for identifying $\rho_s$ in the thermodynamic limit. As explicitly computed in Appendix (\ref{AppA}), it is given by 
\begin{equation}\label{overlapformulaone}
S_{\rho} =\exp\left(-\frac{L}{2}\int_ 0^\infty d \lambda \rho(\lambda +k) \ln \left[\frac{\lambda^2}{c^2}\left(\frac{\lambda^2}{c^2}+\frac{1}{4}\right)\right]-\frac{Ln}{2}\left(\ln\frac{c}{n}+1\right) \right).
\end{equation} 
Comparing it to the overlap between a Bethe state and a non-rotating BEC  \cite{PhysRevA.89.033601}, i.e. (\ref{IS}) with $k=0$, $S_{\rho}$ has the same expression, with the only difference that the root distribution $\rho(\lambda)$ is shifted of $k$.

Let us now impose $\left.\frac{\delta S^{QA}[\rho]}{\delta \rho}\right|_{\rho_s}=0$.
Given (\ref{YYeq}), (\ref{QAeq}) and (\ref{overlapformulaone}) the $\rho$-dependent part  of the quench action per unit length reads
\begin{eqnarray}
\frac{S^{QA}[\rho]}{L}&=&\int _0^ \infty d\lambda  \left\{ \rho(\lambda +k)\ln\left[\frac{\lambda^2}{c^2}\left(\frac{1}{4}+\frac{\lambda^2}{c^2}\right)\right]-\left[(\rho(\lambda +k)+\right.\right.\nonumber\\
&&\left. \left.+\rho^h(\lambda +k))\ln(\rho(\lambda +k)+\rho^h(\lambda +k))-\rho(\lambda +k) \ln\rho(\lambda +k)-\rho^h(\lambda +k) \ln\rho^h(\lambda +k)\right]\right\}.\nonumber
\end{eqnarray}
We have to impose the normalization condition  $\int_{-\infty}^{+\infty}d \lambda \,\rho(\lambda)=n$; this can be done modifying our functional measure by adding a Lagrange multiplier  $\mu$ \cite{1993qism.book.....K} in the following way
\begin{equation}\label{hdeff}
\int {\cal D}[\rho]e^{-S^{QA}[\rho]}\to \int_{-i \infty}^{i \infty}d \mu \int {\cal D}[\rho]\exp\left[-S^{QA}[\rho]-\frac{L \mu}{2}\left(n-\int_{-\infty}^{\infty}d \lambda \, \rho(\lambda)\right)\right].
\end{equation}
Imposing that the variation of the quench action  under an infinitesimal transformation of the Bethe roots be vanishing we get
\begin{equation}\label{hintro}
0=\frac{\delta S^{QA}[\rho]}{L~ \delta \rho}=-\mu+\ln\left(\frac{\rho(\lambda +k)}{\rho^h(\lambda +k)}\right)+\ln\left[\frac{\lambda^2}{c^2}\left(\frac{1}{4}+\frac{\lambda^2}{c^2}\right)\right] -\frac{K}{2\pi}*\ln\left(1+\frac{\rho(\lambda +k)}{\rho^h(\lambda +k)}\right).
\end{equation}
This is an  equation for the root density  $\rho\equiv\rho(\lambda, c, k, \mu)$. In particular, $*$ denotes the convolution product defined as $f *g(k)\equiv \int_{-\infty}^{\infty} d k' f(k-k') g(k')$ and  the Bethe equations have already been imposed at the level of the relation between $\delta\rho^h(\lambda)$ and $\delta\rho(\lambda)$.
Switching to dimensionless variables $x\equiv \frac{\lambda}{c}$ and defining $\tau\equiv e^{\mu/2}$ and $a(x, \tau, \alpha)$ as the ratio of the particle to hole density
\begin{equation}\label{eqa}
a(x, \tau, \alpha)\equiv \frac{\rho(\lambda, c, k, \tau)}{\rho_h(\lambda, c, k, \tau)},
\end{equation}
the saddle point condition (\ref{hintro}) can be rewritten as ($ \alpha=\frac{k}{c}, y=x+\alpha$)
\begin{eqnarray}\label{sprot}
\ln(a(x,  \mu, \alpha))=\ln \left(\frac{\tau ^2}{\left(x-\alpha \right)^2 \left(\frac{1}{4}+(x-\alpha)^2\right)}\right)+\int_{-\infty}^{\infty} \frac{d y}{2 \pi} {\cal K}(x-y) \ln (1+a(y,  \mu, \alpha)),
\end{eqnarray}
where  $ {\cal K}(x)\equiv\frac{1}{1+x^2}$. The connection between   $a(x,  \tau, \alpha)$ and the root density is obtained by applying on the previous equation the operator $D_\tau\equiv\frac{\tau}{2}\partial_\tau$
\begin{equation}\label{eq30}
\frac{\tau}{4 \pi}\partial_\tau \ln a(x,  \tau, \alpha)=\frac{1}{2\pi}+ \frac{\tau}{4\pi} \int_{-\infty}^{+\infty} \frac{d y}{2 \pi}{\cal K}(x-y)\frac{\partial_\tau a(y,  \tau, \alpha)}{1+a(y,  \tau, \alpha)}
\end{equation}
and comparing it to the Bethe equation written in dimensionless variables 
\begin{equation}\label{Betheadim}
\tilde{\rho}(x,  \tau, \alpha)+\tilde{\rho}_h(x,  \tau, \alpha)=\frac{1}{2\pi}+\int_{-\infty}^{+\infty}  \frac{d y}{2\pi}\tilde{\rho}(y, \tau, \alpha){\cal K}(x-y),
\end{equation}
where $\tilde{\rho}(x,  \tau, \alpha)\equiv \rho(\lambda=c x-k,c, k, \tau)$.

Eventually we get that the root density  expressed in terms of the rescaled variables, $\tilde \rho(x,  \tau, \alpha)$ has the form
\begin{equation}\label{a-rho-rel}
\tilde{\rho}(x,  \tau, \alpha)=\frac{\tau \partial_\tau a(x,  \tau, \alpha)}{4\pi (1+a(x,  \tau, \alpha))}.
\end{equation}
In conclusion, finding the saddle point distribution $\rho_s$ amounts to solving  the  non-linear integral equation (\ref{sprot}) for the function $a(x, \tau, \alpha)$ and then plugging it into (\ref{a-rho-rel}).

\begin{figure}[!htb] 
\begin{center} 
\includegraphics[width=.49\textwidth]{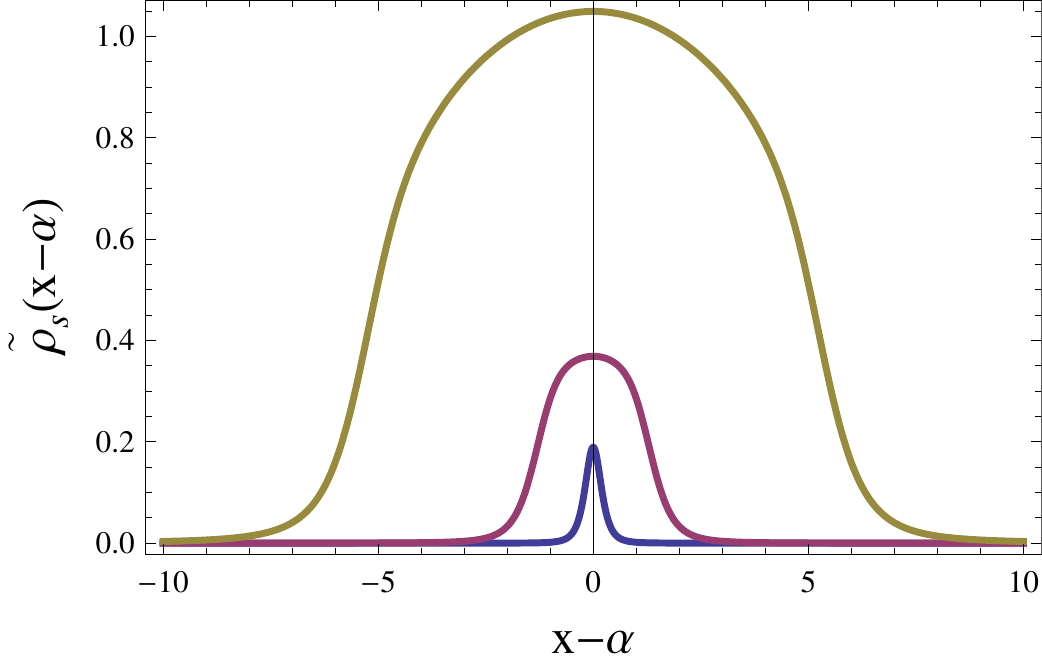} 
\caption{\label{fig:plotsrhorotbec}\small Plot for the saddle point distribution $\tilde\rho_s(x,  \tau, \alpha)$ represented as a function of $x-\alpha$ for $\tau=0.1$ (blue curve),  $\tau=1$ (violet curve),  and  $\tau=10$ (brown curve).}
\end{center} 
\end{figure}

A perturbative study of the solution $\rho_s$ can be used as a guideline towards its full analytic solution. Let us focus on Eq. (\ref{sprot}) in the   $\tau\to 0$ limit. As $\tau\to 0$, for any fixed $x>\tau$,  $\ln \frac{\tau ^2}{(x^2-\alpha^2)^2 ((x-\alpha)^2+\frac{1}{4})}$  is a negative, divergent quantity, so the convolution integral gives a subdominant contribution. Hence the zero-th order for $a$ is
\begin{equation}\label{35}
a^{(0)}(x,  \tau, \alpha)=\frac{\tau^2}{(x-\alpha)^2 ((x-\alpha)^2+1/4)}.
\end{equation}
As it could be argued from the analogy between  (\ref{overlapformulaone}) and the expression of the overlap for the initial BEC state, (\ref{35}) coincides with the zero-th order term of the function $a$ for the  initial BEC state \cite{PhysRevA.89.033601}, upon adding  a constant shift of $\alpha$ to the dimensionless variables. We have checked that this argument  holds also   for higher order terms in $\tau$, hence the full analytic solution for $a(x)$ is
\begin{equation}\label{f11}
a(x,  \tau, \alpha)=\frac{2\pi \tau}{(x-\alpha)\sinh(2\pi(x-\alpha))}I_{1-2 i(x-\alpha)}(4\sqrt{\tau})I_{1+2 i(x-\alpha)}(4\sqrt{\tau}).
\end{equation}
Eq. (\ref{f11}) is the main result of this section. After applying (\ref{a-rho-rel}), the  saddle point  distribution  is derived, yielding
\begin{equation}\label{g11}
\tilde\rho_s(x,  \tau, \alpha)=\tilde\rho^{BEC}_s(x-\alpha,  \tau ).
\end{equation}
Being its exact expression mathematically involved, we prefer showing the plots in Fig. \ref{fig:plotsrhorotbec} for the saddle point distribution as a function of $x-\alpha$, for several values of $\tau$.  As we can see from (\ref{f11}),  $\alpha$ only acts  as a reshift of the centre of the function $a$; the same happens for the distribution $\tilde\rho_s$.

The parameter $\tau$ can be related to the dimensionless interaction constant of the Lieb-Liniger, namely $\gamma$ \cite{PhysRevLett.81.938,1703687,PhysRevLett.91.040403,1367-2630-5-1-379}. Defined as the ratio of the interaction strength $c$ and the particle density $n=N/L$, $\gamma$  is the key-parameter of the Lieb-Liniger model, in terms of which all the physical properties of the system can be described (at zero temperature). Thanks to (\ref{defroot}), it  is related to the root distribution   by
\begin{equation}\label{intrho}
\gamma(\tau, \alpha)\equiv \frac{c}{n}=\frac{1}{\int_{-\infty}^{\infty}dx \tilde\rho_s(x, \tau, \alpha)}.
\end{equation}
 Integrating numerically (\ref{intrho}) for several fixed values of $\tau$, we find that for the specific case of the rotating BEC $\gamma_{BEC}$ is independent of $\alpha$ and is related to $\tau$ as follows
\begin{equation}\label{gammabec}
\gamma_{BEC}(\tau) =\frac{1}{\tau}.
\end{equation}

We would like to highlight that, in this quench protocol from $c=0$ to $c$ finite,  $\gamma$ is also directly a measure of the quench amplitude. We have a small quench for $\gamma\ll1$ \cite{Wadati,WadatiKato,doi:10.1143/JPSJ.54.3727}, an intermediate quench for $\gamma\simeq 1$ and a large quench for $\gamma\gg 1$ \cite{1703687, PhysRevLett.110.245301,2013JSMTE..09..025C,PhysRevA.89.013609}. 

In conclusion we have found that the shape of the root distribution in the stationary state does not change if the initial state, evolved with the Lieb-Liniger, is a non-rotating or rotating BEC. The rotating BEC can be thought of as a state with a steady current, given by the bosons all moving in the same direction with the same value of the momentum. The  effect of this current  is only that of shifting uniformly the whole distribution of the value $k$. We can a posteriori understand it noticing that the problem still remains translationally invariant, for any value of $k$, and  the current does not introduce any additional effect since the interactions between the particles are point-like. As we will see in the next section, the same conclusion does not hold if the bosons have different values of the momentum.


\section{Quench from a state with oppositely moving BECs to the Lieb Liniger}\label{KINO}
In this section we will present the solution of the saddle point distribution for the same quench of the interaction strength from $H_{LL}(c=0)$ to $H_{LL}(c>0)$  on an initial state given by two colliding BECs. 

\begin{figure}[t] 
\begin{center} 
\includegraphics[width=.3\textwidth]{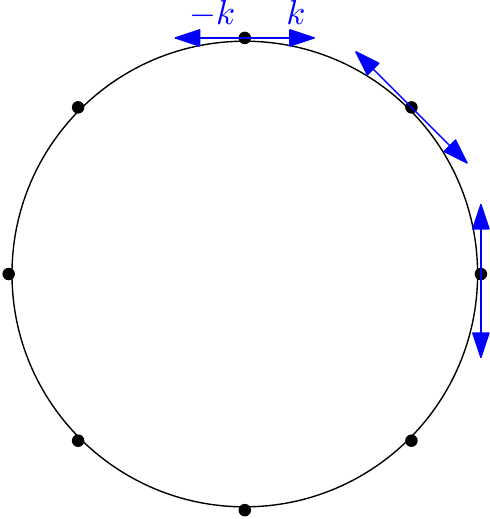}
\caption{\label{fig:Kinofig}\small Schematic representation of the initial state. Each of the $N$ bosons on the ring is in a superposition of states of with definite momentum $\pm k$.}
\end{center} 
\end{figure} 

The initial configuration is   a product state in which each of the $N$ bosons is in a  superposition of right and left moving plane wave states with a definite value of the momentum (see Fig. \ref{fig:Kinofig}), ie.
\begin{equation}
\langle {\bf x}|0^{\leftrightarrows}\rangle_n =\frac{1}{f(k)^N}\langle {\bf x}|0^{\leftrightarrows}\rangle,\quad\quad \langle {\bf x}|0^{\leftrightarrows}\rangle =\prod_{i=1}^N    \frac{(e^{i k x_i}+e^{-ik x_i})}{\sqrt{2L}},
\end{equation}
with the $k$-dependent normalization factor $f(k)=\sqrt{\frac{kL+\sin(kL)}{kL}}$ ensuring  that $\langle {\bf x}|0^{\leftrightarrows}\rangle_n$  is normalized to 1. Using periodic boundary conditions on the initial state, we have that $k= 2\pi n/L$, hence the normalization constant $f(k)$ reduces to $1$.

To obtain the saddle point root distribution, we follow the same path outlined in the previous section. It can be shown (see Appendix  \ref{AppB}) that, in the thermodynamic limit, the leading contribution to the overlap between this initial state and a Bethe state is given by parity invariant Bethe states. The extensive part of the overlap  is  (see Appendix  \ref{AppB} for the detailed computation)
 \begin{equation}\label{one2}
S_{\rho}= \exp\left\{- \frac{L}{2}  \int_0^{\infty} d \lambda \rho(\lambda) \ln \left[  \left(  \frac{\lambda }{c}-\alpha^2\frac{c}{\lambda}\right)^2  \left(\frac{1}{4}+\frac{\lambda^2}{c^2}\right) \right]-\frac{Ln}{2}\left(1+\ln\frac{c}{n}\right)\right\}.
 \end{equation}
We note that, irrespectively of the boundary conditions used, the normalization factor $f(k)$ of the initial state can be dropped in $S_{\rho}$ because it does not contribute to the saddle point distribution, being $\rho$-independent.

Using  (\ref{YYeq}), (\ref{QAeq}) and (\ref{one2}) to construct the quench action, extremizing it with respect to $\rho(\lambda)$ and switching to dimensionless variables $x=\frac{\lambda}{c}$ yields the following non-linear integral equation
\begin{eqnarray}\label{sprot2}
\ln(a(x, \tau, \alpha))=\ln \frac{\tau ^2 x^2}{(x^2-\alpha^2)^2 (x^2+\frac{1}{4})}+\int_{-\infty}^{\infty} \frac{d y}{2 \pi} {\cal K}(x-y) \ln (1+a(y, \tau, \alpha)),
\end{eqnarray}
with $a(x, \tau, \alpha)$ and $\tau$ defined as in Sec. \ref{rotBEC}.

 Eq. (\ref{sprot2}) differs from (\ref{eq30}), obtained for the initial rotating BEC, only in the contribution of the overlap. Once it  is solved for  $a(x, \tau, \alpha)$, the  root distribution  can be derived using the definition of $a(x, \tau, \alpha)$ and the dimensionless Bethe root equation (\ref{Betheadim})  written in terms of $\tilde\rho(x, \tau, \alpha)\equiv\rho(\lambda=c x, c,k, \tau )$.
Since the analytic solution is highly non trivial for generic values of $\tau$ and $\alpha$, we will discuss the  solution obtained from the numerical integration of Eq. (\ref{sprot2}),   for  various values of $\alpha$ and $\tau$.

\begin{figure}[t] 
\begin{center} 
\includegraphics[width=.45\textwidth]{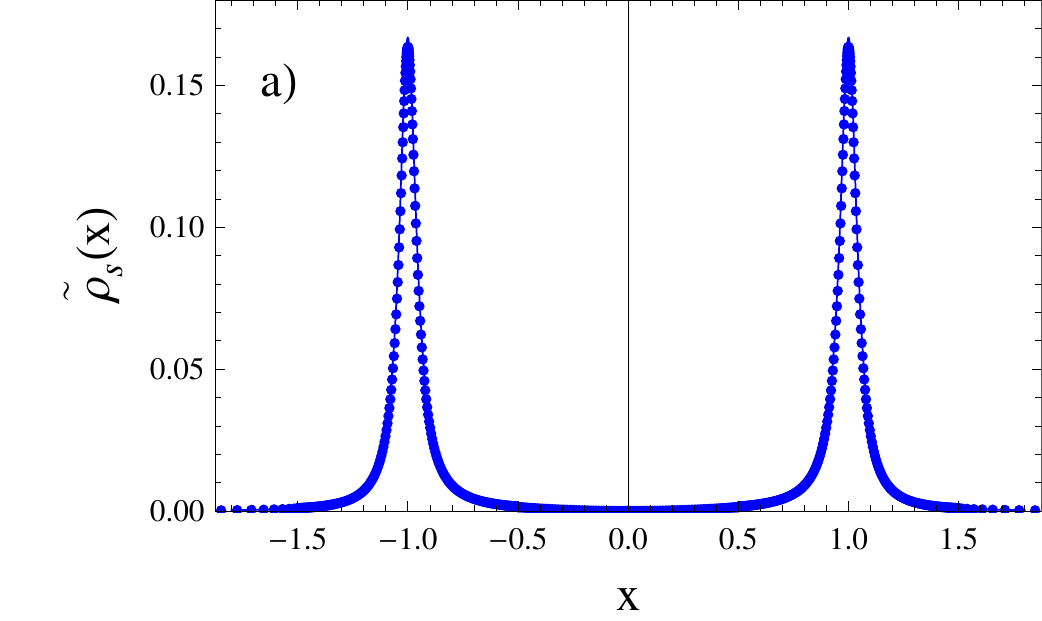}~~~~~~~~~~~~~ \includegraphics[width=.45\textwidth]{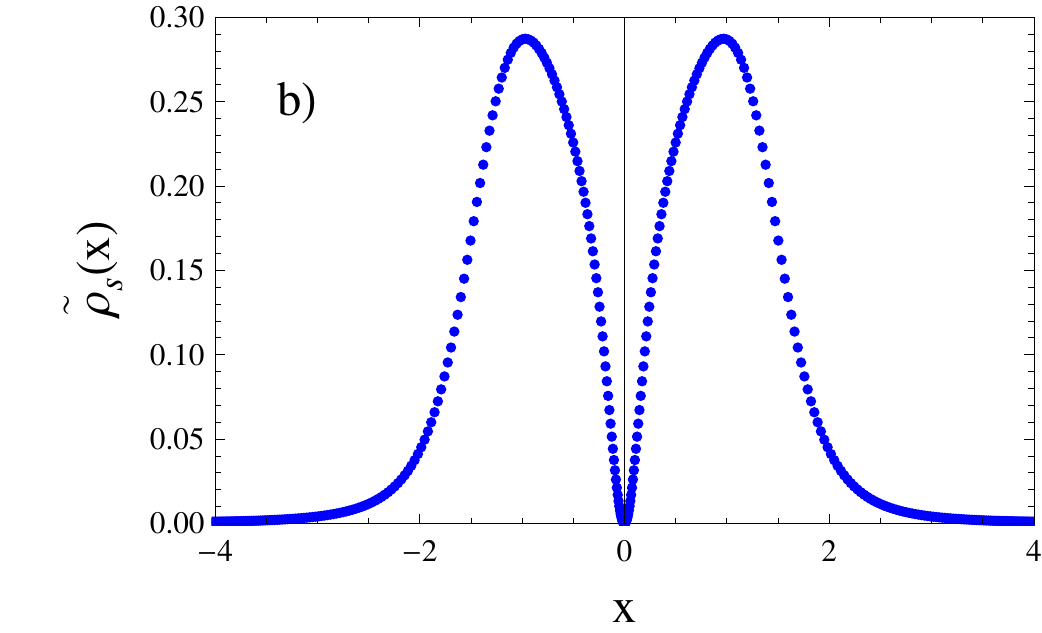} \includegraphics[width=.45\textwidth]{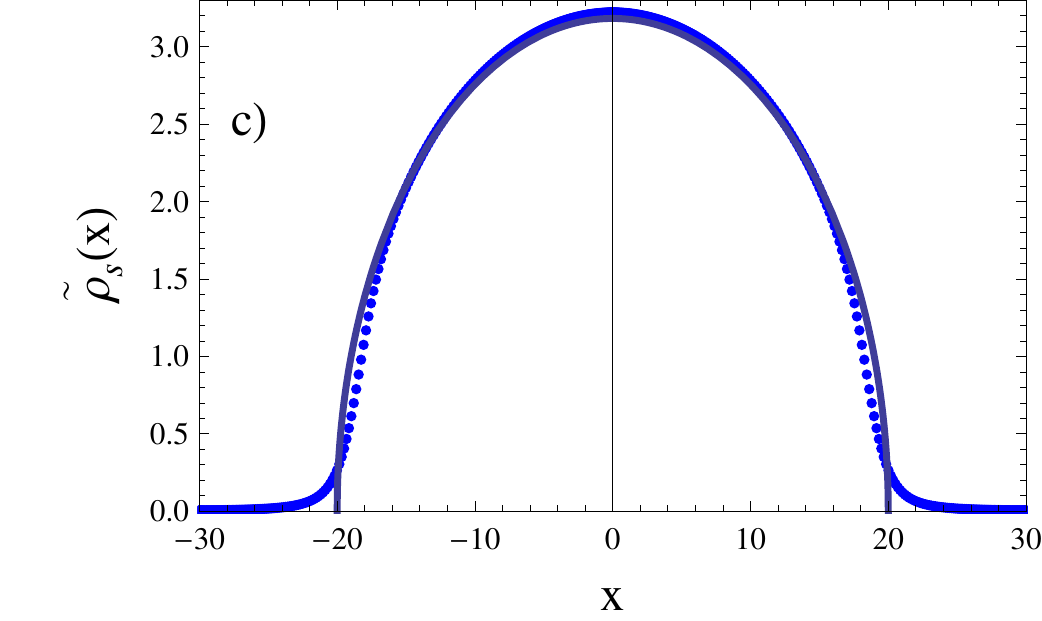}
\caption{\label{fig:PLOTRHO}\small Plots  for $\tilde \rho_s(x, \tau, \alpha)$ as a function of $x$ for $\alpha=1$ and $\tau=0.1$ (a), $\tau=1$ (b) and $\tau=100$ (c).  (a) is the representative case for the large quench regime ($\gamma\gg 1$),  (b) for the intermediate quench ($\gamma \simeq 1$) and  (c) for the small quench regime ($\gamma \ll 1$). Points represent the numerical solution, the continuous line represents the approximate solution. }
\end{center} 
\end{figure}

\begin{center} 
\begin{figure}[t] 
\centering{\includegraphics[width=.45\textwidth]{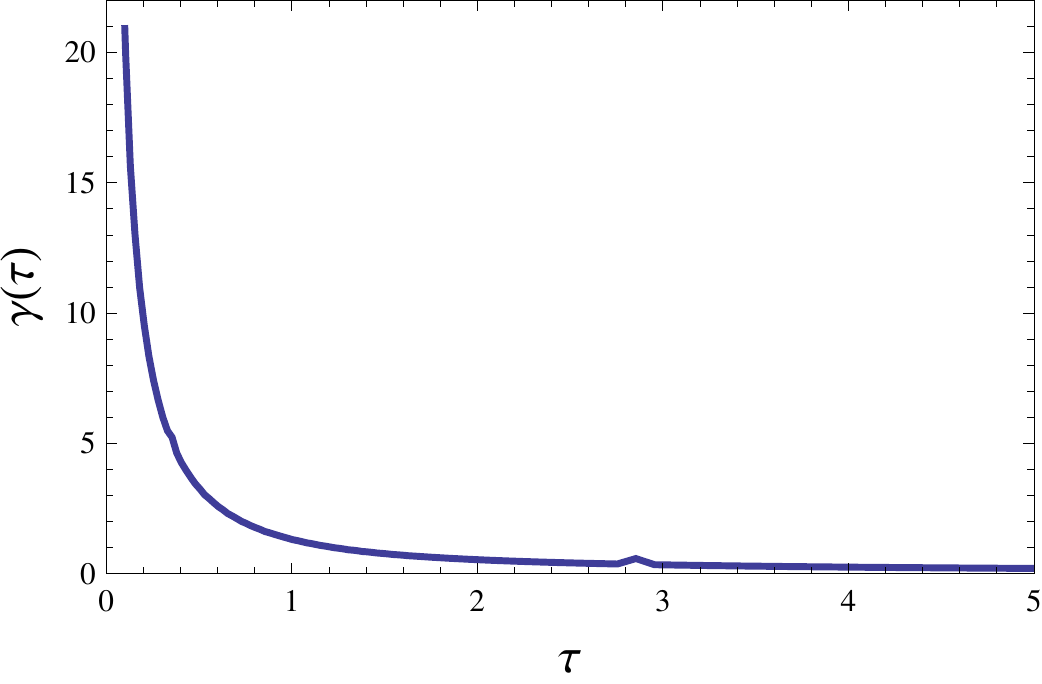}}
\caption{\label{fig:tabe} 
Dependence of $\gamma$ on $\tau$  for $\alpha=1$. Even for non vanishing values of $\alpha$, the large(small) quench regime  corresponds to small(large) values of $\tau$.}
\end{figure} 
\end{center}

In Fig. \ref{fig:PLOTRHO} we show the numerical plots  for the saddle point distribution for $\alpha=1$ as a function of $x$, obtained for $\tau=0.1,1,100$.  Computing $\gamma$ from (\ref{intrho}) for each of the saddle point distributions, we see that these three plots correspond to the three possible quench regimes: Fig. \ref{fig:PLOTRHO}(a) represents a large quench ($\gamma\gg 1$),  Fig. \ref{fig:PLOTRHO}(b) an intermediate quench ($\gamma \simeq 1$) and Fig. \ref{fig:PLOTRHO}(c) a  small quench ($\gamma \ll 1$). Differently from the rotating BEC, in this case  $\gamma$ depends  on both $\alpha$ and $\tau$ but there is still the correspondence between small(large) values of $\tau$ and large(small) values of $\gamma$. As an example, we show in Fig.  \ref{fig:tabe} the  plot  of $\gamma$ as a function of $\tau$ in the case $\alpha=1$.

Computing $\gamma$ for several values of $\alpha, \tau$ and imposing it to be of order one, we see that the separation between small/large quenches sits approximately at $\tau\simeq \mathrm{max}\{\alpha,1\}$.  The classification in terms of the parameter $\gamma$ turns out to be useful, in fact  the saddle point distributions obtained for different values of $\alpha$ and $\tau$ in a specific $\gamma$ regime show the same typical features summarized in the three plots of Fig \ref{fig:PLOTRHO}. For this reason we will only show the plots obtained for the value $\alpha=1$. 

\begin{figure}[!htb] 
\begin{center} 
\includegraphics[width=.45\textwidth]{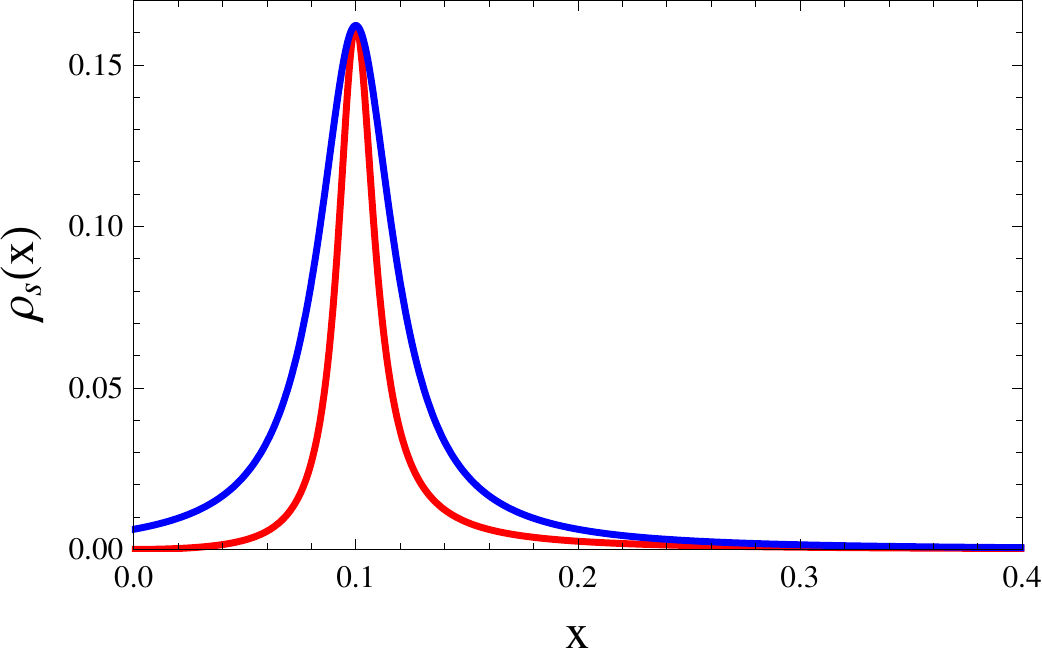}~~~~~~~~~~~~~\includegraphics[width=.45\textwidth]{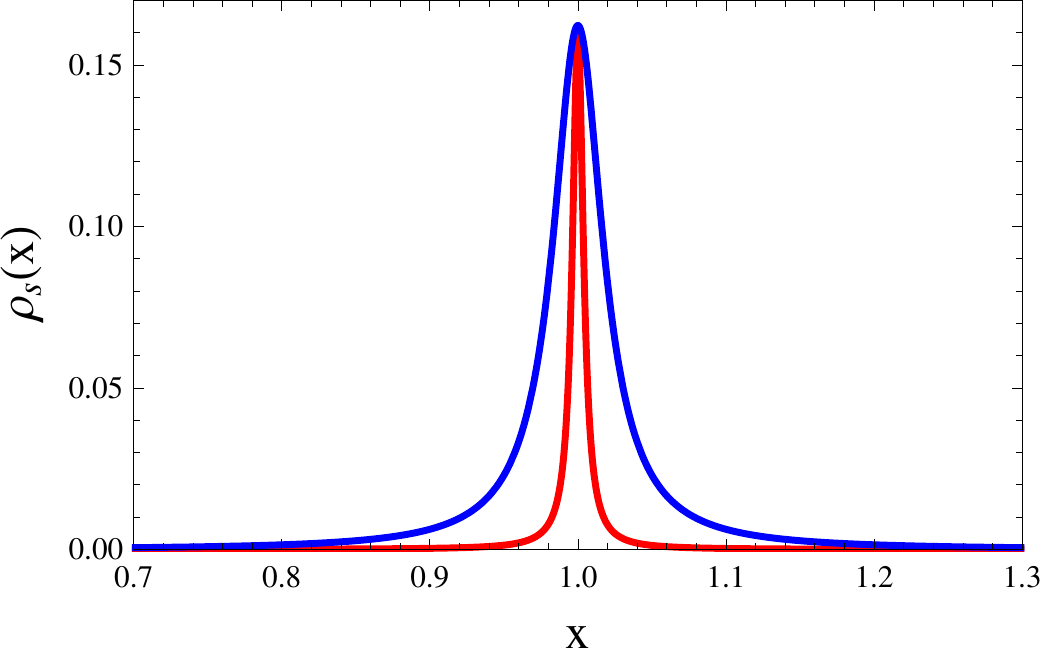} \includegraphics[width=.45\textwidth]{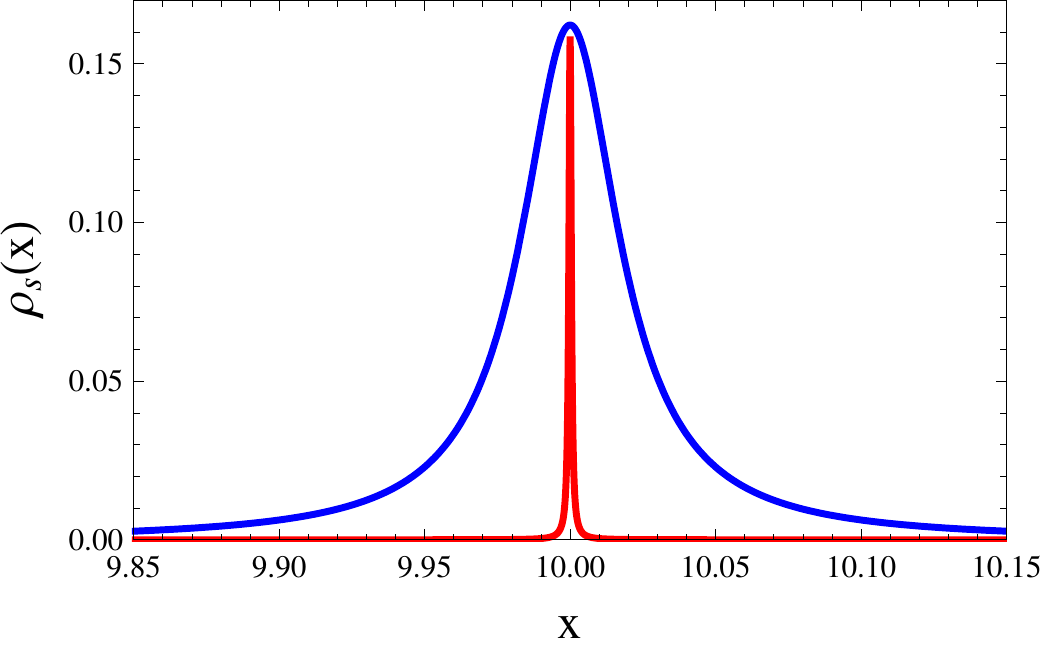}
\caption{\label{fig:Compa} 
Plots for for $\tilde \rho_s^{(1)}(x, \tau, \alpha)$ (red curve)  versus $\tilde\rho^{BEC}_s(x-\alpha, \tau)$ (blue curve) for $\tau=0.01$ and $\alpha=0.1$ (Up Left), $\alpha=1$ (Up Right) and $\alpha=10$ (Down). Only the region of $x>0$ is shown.}
\end{center} 
\end{figure}

Let us now focus on the analysis of the two opposite regime of large and small quenches, where approximate solutions were found, and comment Fig. \ref{fig:PLOTRHO}. For the large quench case, we can use a perturbative expansion in the small parameter $\tau$ since the large quench regime $\gamma\gg 1$ corresponds to $\tau\ll 1$;  up to order  $\tau^3$, the solution is
\begin{equation}\label{pert_res}
\tilde\rho_s^{(1)}(x, \tau, \alpha)=\frac{x^2 \tau ^2\left[6 \tau (1+ x^2+\alpha^2)+\sqrt{1+4 \alpha^2} (1+(x-\alpha)^2)(1+(x+\alpha)^2)\right]}{2 \pi \left[\sqrt{1+4 \alpha^2} (1+(x-\alpha)^2)(1+(x+\alpha)^2) \left(x^2\tau^2 +(x^2-\alpha^2)^2(x^2+\frac{1}{4})\right)+4 x^2\tau^3(1+x^2+\alpha ^2) \right]}.
\end{equation}
It is represented by the continuous line in Fig. \ref{fig:PLOTRHO}(a), which agrees extremely well with the numerical plots  and  is derived in Appendix \ref{AppC}.   The behaviour for large $x$, for $\tau$ small but finite, is given by 
\begin{equation}
\tilde\rho^{(1)}_s(x, \tau, \alpha)\simeq \frac{\tau^2}{2 \pi x^4}+\frac{\tau^2(-1+8 \alpha^2+\frac{24 \tau}{\sqrt{1+4 \alpha^2}})}{8 \pi x^6}+\frac{\tau^2(1-8 \alpha^2+48 \alpha^4+120 \frac{(4 \alpha^2-1) \tau}{\sqrt{1+4 \alpha^2}})}{32 \pi x^6}+\dots,
\end{equation}
whose  leading order is $\frac{1}{x^4}$, the same as for the case of the initial BEC state. To appreciate the difference between the root distribution of the state with oppositely moving BECs and the rotating BEC in the large quench limit we can compare $\tilde\rho^{BEC}_s(x-\alpha, \tau)$ with $\tilde\rho_s ^{(1)}(x, \tau, \alpha)$ (see Fig. \ref{fig:Compa}). The difference between the two plots increases with increasing $\alpha$; it is clear that the saddle point distribution for the  oppositely moving BECs  can not be obtained trivially from   reshifting  the root density distribution of the BEC case.

Let us now focus on  the small quench limit. In this regime, $\tau> \mathrm{max}\{\alpha,1\}$, so  $\alpha$ is negligible with respect to $\tau$; this explains why the root distribution in Fig. \ref{fig:PLOTRHO}(c), for a given value of $\tau$,  is the same for  any values of $\alpha$ ranging over three order of magnitudes, from $0.01$ to $10$. In Fig. 4(c), in addition to the root distribution in the small quench limit, we have also plotted the Wigner-semicircular law (represented by the continuous line)
\begin{equation}\label{WSL}
\tilde \rho_W(x, \gamma)=\frac{1}{\pi \sqrt{\gamma}}\sqrt{1-\frac{x^2 \gamma}{4}}.
\end{equation}
Indeed in \cite{PhysRevA.89.033601}, for the initial BEC state, it was rigorously proved that  the Wigner-semicircular law describes  the stationary root distribution in the case of a small quench (more precisely  in the limit $c\to 0$ with $n$ fixed). Interestingly enough,  (\ref{WSL}) also represents  the leading term of the ground state root distribution of the Lieb-Liniger model in the low-$\gamma$ expansion \cite{PhysRevA.4.386}, i.e. of the ground state of the post-quenched Hamiltonian. Actually, as can be seen in Fig. 4(c), we note that also starting from two counter propagating BECs the stationary root distribution turns out to be very close to the Wigner law, apart from a tail for  $|x|> \frac{1}{\pi \sqrt{\gamma}}$, probably  due to initial excitations.

\section{Conclusions}\label{concl}

In this paper we have studied in detail the asymptotic state reached after a quench to arbitrary values of the interaction strength of the Lieb-Liniger model, from two experimentally relevant initial configurations. In the first case, when considering  bosons forming a rotating BEC state, the steady state root distribution was found to be the same as when the bosons are initially in a BEC configuration, simply shifted  by the momentum of each boson. In the second case, when the initial state can be represented as two oppositely moving  BECs, we showed that the non vanishing momentum of the bosons radically modifies the stationary root distribution. We were able to identify three regimes for the saddle point root distribution, characterized by the parameter $\gamma$ (large, intermediate and small quench regime) and  to find approximate solutions in both the small and large quench limit that fit very well with the numerical data. 

 It would be very interesting to analyze the dependence on the quench amplitude $\gamma$ of the local two and three point function of the stationary state. Such analysis could allow us to compare  with the well known results for the correlation functions in  highly excited thermal states of the Lieb-Liniger \cite{PhysRevLett.91.040403,1367-2630-5-1-379,PhysRevLett.107.230405}.


\subsubsection*{Acknowledgements}  
I acknowledge support and hospitality from SISSA and the Max Planck Institute for the Physics of Complex Systems, Dresden, where this work was completed. I would like to thank Pasquale Calabrese for useful comments and for reading the final version of the manuscript. 

\appendix


\section{Overlap between a Bethe eigenstate and the rotating BEC}\label{AppA}
In this appendix we give the detailed derivation of formula (\ref{overlapformulaone}), which is the extensive part of the overlap between the rotating BEC and a generic eigenstate of $H_{LL}(c>0)$, in the thermodynamic limit.
\subsubsection*{Overlap for generic $N$, $L$} 
Due to momentum conservation, the overlap between a Bethe state and the rotating BEC  is non-vanishing if the sum of the rapidities of the Bethe state is equal to the total momentum of the initial state, $N k$. This class of states includes as a special case the Bethe states with rapidities symmetrically distributed  around $k$. Nevertheless we can safely consider only this subset of states because the Bethe equations are not mutually consistent for Bethe states that are not parity invariant (with respect to $k$).

To obtain the overlap, let us first explicitly  write the generic normalized (subscript $_n$) Bethe eigenstate. It is given by (\ref{HLL}) divided by its norm, the Gaudin determinant \cite{1993qism.book.....K}
\begin{equation}
\langle\boldsymbol{x}|\boldsymbol{\lambda}\rangle_n \equiv  \psi( {\bf x}|{\boldsymbol \lambda})_n=\frac{\psi( {\bf x}|{\boldsymbol \lambda})}{\sqrt{\mathrm{det} G}},
\end{equation} 
where $G$ is a $N \times N$ matrix with elements
\begin{equation}
{G}_{jk}=\frac{\partial ^2 A_{YY}(\boldsymbol{\lambda})}{\partial \lambda_j \partial \lambda_k}.
\end{equation}
 $A_{YY}$ is the Yang-Yang action for the Lieb-Liniger model \cite{1664947}, defined as
\begin{equation}
A_{YY}(\boldsymbol{\lambda})\equiv\frac{L}{2}\sum_{j=1}^N \lambda _j^2+\sum_{j<l=1}^N\Phi (\lambda_j -\lambda_k)-2 \pi I_j \lambda_j,
\end{equation}
 with $\Phi(\lambda)\equiv \int _0 ^\lambda d \lambda'\, 2 \mathrm{arctan} (\frac{\lambda'}{c})$. Since the Hessian of the Yang-Yang action depends only on the difference between rapidities,  which are symmetric with respect to $k$,  the Gaudin determinant is independent of $k$ and has the same expression as for the initial (non-rotating) BEC state, i.e.
 \begin{equation}
 \mathrm{det} G=\mathrm{det}_{j,k=1}^N\left[\delta_{jl}\left(L+\sum_{l=1}^{N} K(\lambda _l- \lambda _k)\right)-K(\lambda _j- \lambda _k)\right].
\end{equation}
The overlap can thus be written as
\begin{eqnarray}\label{uno}
&&\langle  0^\rightarrow|\boldsymbol{\lambda}\rangle_n =\int_ 0^L d x_1\cdots \int_ 0^L d x_N  \frac{1}{L^{N/2}}\prod_{i=1}^N e^{-i k x_i}\frac{\psi(\boldsymbol{ x}|{\boldsymbol \lambda})}{\sqrt{\mathrm{det} G}}= \frac{I^\rightarrow (L) N!}{L^{N/2}\sqrt{\mathrm{det} G}},
\end{eqnarray}
where we have defined 
\begin{equation} 
I^\rightarrow (L)\equiv  \int_{\boldsymbol{1}} d^N x \,\psi(\boldsymbol{ x}|{\boldsymbol \lambda})  \prod_{i=1}^N e^{-i k x_i}
\end{equation}
and we recall that $\boldsymbol{1}=\{0\leq x_1<x_2< \dots <x_N\leq L\}$. To compute $I^\rightarrow (L)$  let us explicitly write the  expression for  the not-normalized eigenfunction 
 \begin{equation} \label{ext_un_wf}
\psi( \boldsymbol{ x}|{\boldsymbol \lambda})=  \sum_{P\in S^N} \left[\prod_{j>l=1}^N  \frac{ (\lambda_j-\lambda_l)}{\sqrt{N!((\lambda_j-\lambda_l)^2+c^2)}} e^{i\lambda_{P_l} x_l} \left(1-ic \frac{sgn(x_j-x_l)}{\lambda_{P_j}-\lambda_{P_l}}\right)  \right];
 \end{equation}
hence we get 
\begin{eqnarray}\label{as}
I^\rightarrow (L)&=&\sum_{P\in S^N} \left[\prod_{j>l=1}^N \frac{ (\lambda_j-\lambda_l)}{\sqrt{N!((\lambda_j-\lambda_l)^2+c^2)}} e^{i\lambda'_{P_l} x_l} \left(1-ic \frac{sgn(x_j-x_l)}{\lambda'_{P_j}-\lambda'_{P_l}}\right)  \right],
\end{eqnarray}
where $\lambda'_l \equiv\lambda_l-k$. This is exactly of the same form as the one computed for the initial BEC case \cite{PhysRevA.89.033601}, provided we substitute the rapidities $\lambda$ with the distance of the rapidities from the centre $k$ of the distribution $\lambda'$.
  By exploiting the properties of the Laplace transform of the overlap \cite{2012JSMTE..06..001L,1742-5468-2014-5-P05004}, (\ref{as}) can thus be rewritten generalizing to the present case the expression for the BEC of  \cite{PhysRevA.89.033601}
\begin{equation}\label{afterLaplaceT} 
I^\rightarrow (L)=   \sum_{P\in S^N}\left[\prod_{j>l=1}^N  \frac{(\lambda'_j-\lambda'_l)}{\sqrt{N!((\lambda'_j-\lambda'_l)^2+c^2)}}\left(1-\frac{ic}{\lambda'_{P_j}-\lambda'_{P_l}}\right)\sum_{res}\mathrm{Res}_s\left(\frac{e^{sL}}{s}\prod_{j=1}^N\frac{1}{s-i\sum_{l=j}^N\lambda'_{P_l}}\right)\right],
\end{equation}
where $\mathrm{Res}_s$ indicates to take the residue with respect to the $s$-variable. 
\subsubsection*{Extensive part of the overlap in the thermodynamic limit}
Deriving the thermodynamic limit of the overlap is a non trivial task (see for example \cite{2008arXiv0804.2543B}) but, for the computation of the quench action (\ref{QAeq}), we just need the extensive part of it. In \cite{PhysRevA.89.033601} the full expression of the overlap in the thermodynamic limit  was analytically computed for the initial BEC state and its extensive part  turned out to coincide with the leading contribution of its zero-density limit.  In the present case, we first compute the zero density limit of the overlap and then we check that it  coincides with the leading order in $L$ of the overlap computed in the finite density case $N=2$, $L$ finite. Based on the analogy with the exact calculation in the BEC case, this gives us the sought expression for the extensive part of the overlap in the thermodynamic limit.

To compute the zero-density limit of (\ref{afterLaplaceT}),  we have to isolate the leading term in $L$ when the  number of particles $N$ is kept finite. In this case the computation is simplified with respect to the thermodynamic limit because  the rapidities are not quantized, hence we do not have to enforce any Bethe equations. Since, for any $N$, $L$, the overlap with the rotating BEC has the same expression (in terms of $\lambda'$) as the overlap of the BEC, this  holds also in the zero-density limit; thus we can straightforwardly consider the expression obtained in \cite{PhysRevA.89.033601} and adapt it to our rotating case
 \begin{eqnarray}\label{overlapzerodensity}
&&I^\rightarrow (L)=\frac{(-L/c)^{N/2}}{\sqrt{N!}}\frac{1}{\prod_{j=1}^{N/2}\frac{\lambda_j-k}{c}\sqrt{\frac{1}{4}+\frac{(\lambda_j-k)^2}{c^2}}}+ {\cal O}(1/L).
\end{eqnarray}
Since the Gaudin determinant, to the leading order in $L$, is of the form
\begin{equation}
\mathrm{det}G\simeq L^N+{\cal O}(L^{N-1}),
\end{equation}
the zero-density limit of (\ref{uno}) becomes
\begin{equation}
\langle  0^\rightarrow|{\boldsymbol \lambda}\rangle_n = \frac{\sqrt{(cL)^{-N}N!}}{\prod_{j=1}^{N/2}\frac{\lambda_j-k}{c}\sqrt{\frac{1}{4}+\frac{(\lambda_j-k)^2}{c^2}}}(1+O(1/L)).
\end{equation}
The expansion in $1/L$ can be rewritten as an expansion in $n$, $n$ being the density of particles $n=N/L$, which reads 
\begin{equation}\label{jjj}
\langle  0^\rightarrow|{\boldsymbol \lambda} \rangle_n =\exp\left(-\frac{L}{2}\int_ 0^\infty d \lambda \rho(\lambda +k) \ln \left[\frac{\lambda^2}{c^2}\left(\frac{\lambda^2}{c^2}+\frac{1}{4}\right)\right]-\frac{Ln}{2}\left(\ln\frac{c}{n}+1\right) \right)(1+{\cal O}(n)).
\end{equation}
The only difference from the non rotating case is that here $\rho$ is shifted of $k$. 

Let us  compute the overlap in a finite density case, i.e.  $N=2$ with $L$ finite. Working out calculations from (\ref{afterLaplaceT}) with the Bethe state $|\lambda_1,\lambda_2\rangle=|k+\beta, k-\beta\rangle$ we get
\begin{equation}
I^\rightarrow_{N=2} (L)= \frac{2\beta}{\sqrt{2(4\beta^2+c^2)}}\left[e^{iL\beta}\frac{2\beta-ic}{2\beta^3}+e^{-iL\beta}\frac{2\beta+ic}{2\beta^3}-\frac{2+cL}{\beta^2}\right].
\end{equation}
Enforcing the Bethe equation
\begin{equation}
e^{iL \beta}=\frac{2\beta+ic}{2\beta-ic}e^{-iLk},
\end{equation}
 the overlap reads
\begin{equation}\label{overlap2}
\langle  0^\rightarrow|\{k+\beta , k-\beta\}\rangle_n =\frac{\sqrt{2}}{L \beta\sqrt{1/4+(\beta/c)^2}} \left[1+\frac{1}{L}\left(\frac{4}{c}\sin^2\frac{kL}{2}-\frac{1}{\beta}\sin(kL)\right)\right].
\end{equation}
The k-dependent term in (\ref{overlap2}) is subleading in $L$. Given the  non trivial form of  the overlap already for the finite size case with $N=2$, we can extrapolate that any $k$-dependent term will give subleading contributions for any finite size case. The extensive part of the overlap in the thermodynamic limit is thus given by  (\ref{jjj})
\begin{equation}\label{overlapformula}
S_{\rho} =\exp\left(-\frac{L}{2}\int_ 0^\infty d \lambda \rho(\lambda +k) \ln \left[\frac{\lambda^2}{c^2}\left(\frac{\lambda^2}{c^2}+\frac{1}{4}\right)\right]-\frac{Ln}{2}\left(\ln\frac{c}{n}+1\right) \right).
\end{equation} 

\section{Overlap between a Bethe eigenstate and a state with oppositely moving BECs}\label{AppB}

In this appendix we derive formula (\ref{one2}). We first compute the expression for the overlap between a state with oppositely moving BECs and a generic Bethe state for generic $N$, $L$. Then we derive its zero-density limit which we identify with the extensive part of the overlap in  the thermodynamic limit similarly to what done in  Appendix \ref{AppA}.

\subsubsection*{Overlap for generic $N$, $L$}

 The overlap is \footnote{ The pedex ${\,}_n$ refers to the normalization of the Bethe state $|\boldsymbol\lambda\rangle $.} 
\begin{eqnarray}
&&\langle  0^\leftrightarrows|\boldsymbol \lambda\rangle _n=\int_{\boldsymbol 1} d^N x \frac{1}{(2L)^{N/2}}\prod_{i=1}^N \left(e^{-i k x_i}+e^{i k x_i}\right)\frac{\psi(\boldsymbol{x}|{\boldsymbol \lambda})N!}{\sqrt{\mathrm{det} G}}.\nonumber\\
\end{eqnarray}
Defining
\begin{equation}
I^\leftrightarrows (L)\equiv  \int_{\boldsymbol 1} d^N x\ \psi(\boldsymbol{ x}|{\boldsymbol \lambda})  \prod_{i=1}^N \left(e^{-i k x_i}+e^{i k x_i}\right),
\end{equation}
the expression for the overlap is
\begin{equation}\label{exp_overlap}
\langle  0^\leftrightarrows|\boldsymbol \lambda\rangle _n=  \frac{I^\leftrightarrows (L) \quad N!}{(2L)^{N/2}\sqrt{\mathrm{det}G}}.
\end{equation}

 Let us compute $I^\leftrightarrows (L)$. By inserting the expression for the Bethe wavefunction (\ref{ext_un_wf}) in (\ref{exp_overlap}) we get
\begin{equation}
I^\leftrightarrows (L)=  \int_{\boldsymbol 1} d^N x \sum_{P\in S^N} \left[ \prod_{j>l=1}^N  \frac{(\lambda_j-\lambda_l)}{\sqrt{N!((\lambda_j-\lambda_l)^2+c^2)}}\left(e^{i(\lambda_{P_l}-k) x_l}+ e^{i(\lambda_{P_l}+k) x_l}\right) \left(1- \frac{ic}{\lambda_{P_j}-\lambda_{P_l}}\right)  \right],
\end{equation}
which can be written in a more compact form as
\begin{equation}\label{a22}
I^\leftrightarrows (L)=  \int_{\boldsymbol 1} d^N x \sum_{P\in S^N} \left[\sum_{{\boldsymbol k}} {\cal A}_{\boldsymbol\lambda_P, \boldsymbol k}  \prod_{j>l=1}^N \frac{(\lambda_j-\lambda_l)}{\sqrt{N!((\lambda_j-\lambda_l)^2+c^2)}}\left(1- \frac{ic}{\lambda_{P_j}-\lambda_{P_l}} \right) \right].\nonumber\\
\end{equation}
In the previous equation we have defined ${\boldsymbol\lambda}_P=\{ \lambda_{P_1},  \lambda_{P_2},\dots, \lambda_{P_N} \}$ as the set of   rapidities  associated to the particles at positions $\boldsymbol x=\{x_1, x_2,\dots , x_N\}$ within the permutation $P$ and $\boldsymbol k=\{ k_1,  k_2, \dots, k_N\}$ as the set of momenta associated to the particles at positions $\boldsymbol x$, with $k_i=\pm k$, $ i=1\dots N$; lastly we have defined
\begin{equation}
{\cal A}_{\boldsymbol\lambda_P, \boldsymbol k} \equiv \exp\{i \boldsymbol x \cdot (\boldsymbol\lambda_P + \boldsymbol k)\}.
\end{equation} 

Differently from the rotating case, a state with two oppositely moving BECs  is not an eigenstate of the momentum operator hence, in principle, it could have non vanishing overlap with non parity symmetric Bethe states. Nevertheless it is possible to show that in the thermodynamic limit the leading contribution to the overlap is obtained just by considering the parity symmetric Bethe states.
First of all, let us stress again that, since the extensive part of the overlap in the thermodynamic limit coincides with the extensive part of the overlap in the zero density limit, it is enough to prove the previous statement in the zero density limit. We have explicitly computed the overlap for a generic $N=2$ Bethe state, which is an already complicated expression. Considering opposite rapidities, which is only possible realization of parity symmetric state for $N=2$, we get
\begin{equation}\label{11}
I^\leftrightarrows (L)(\{  \lambda_1,\lambda_2\} )=\frac{16 i c k^2 \left(-k^2+\lambda_1 ^2+\lambda_2 ^2+\lambda_1\lambda_2\right) \left(-1+e^{i L (\lambda_1+\lambda_2)}\right)}{(\lambda_1+\lambda_2) (\lambda_1-k) (k+\lambda_1) (\lambda_2-k) (k+\lambda_2) (-2 k+\lambda_1+\lambda_2) (2 k+\lambda_1+\lambda_2)}.
\end{equation}
In the $L\to \infty$ limit, considering that $\lambda_1+\lambda_2=0$ and enforcing the Bethe equations, (\ref{11}) is extensive in the system's size
\begin{equation}
I^\leftrightarrows (L)(\{ \lambda_1, -\lambda_1\}) =-\frac{4 c L}{k^2-\lambda_1^2}.
\end{equation}

We explicitly checked that the non parity invariant states give a subleading contribution to the overlap, which can be neglected in the thermodynamic limit. Hence, we can safely compute (\ref{a22}) only for  parity invariant Bethe states.

Let us now go back to (\ref{a22}). If we consider a single term in the summation over $\boldsymbol k$,  (\ref{a22}) represents the overlap between a Bethe state with rapidities ${\boldsymbol\lambda}_P$ and a state where the bosons have momenta ${\boldsymbol k}$.   For $N=2$, given a Bethe state $\{\beta,-\beta\}$, the only relevant combinations are $\{\beta+k,-\beta-k\}, \{\beta-k,-\beta+k\}$. We will label $Q$ as an element of the permutation group acting on a parity symmetric $\boldsymbol k$ state and we will denote  ${\boldsymbol\lambda}_P^Q\equiv \boldsymbol \lambda_P+  \boldsymbol k^Q $. Exploiting the properties of the Laplace transform \cite{PhysRevA.89.033601,2012JSMTE..06..001L,1742-5468-2014-5-P05004}, (\ref{a22}) can be rewritten as
\begin{equation}\label{afterLaplaceTlr}
I^\leftrightarrows(L)=    \sum_{P, Q }\left[\prod_{j>l=1}^N \frac{ (\lambda_j-\lambda_l)}{\sqrt{N!((\lambda_j-\lambda_l)^2+c^2)}} \left(1-\frac{ic}{\lambda_{P_j}-\lambda_{P_l}}\right)\sum_{res}\mathrm{Res}_s\left(\frac{e^{sL}}{s}\prod_{j=1}^N\frac{1}{s-i\sum_{l=j}^N\lambda_{P_l}^Q}\right)\right],
\end{equation} 
where $\mathrm{Res}_s$ indicates to take the residue with respect to the $s$-variable. 
 
 \subsubsection*{Extensive part of the overlap in the thermodynamic limit}
 We will follow the same path outlined in Appendix \ref{AppA}, which consists in deriving first the zero density limit of the overlap and then computing the leading contribution in $L$ in the finite density case.
 
To compute the zero density limit of the overlap, we have to isolate the leading term in $L$ keeping $N$ finite. Let us first consider how this is done for the  $k=0$ case and then discuss the general $k \neq 0$ case.
 
  Setting $k=0$ in (\ref{afterLaplaceTlr}), the highest order in $L$ is carried by the pole in $s=0$; it is of order $\frac{N}{2}+1$ if and only if the permutation acting on a reference configuration $R$ gives as result a  target configuration $T$, such that  $R$ and $T$ are of this form
 \begin{eqnarray}
 &&\quad R: \quad\quad \{\lambda_1,-\lambda_1,\cdots, \lambda_j,-\lambda_j, \cdots,\lambda_k, -\lambda_k,\cdots, \lambda_{N/2},-\lambda_{N/2}\}\nonumber\\
 &&\nonumber\\
 &&\quad T: \quad\quad \{\lambda_1,-\lambda_1,\cdots, \lambda_k,-\lambda_k, \cdots,\lambda_j, -\lambda_j,\cdots, \lambda_{N/2},-\lambda_{N/2}\}\nonumber\\
 &&\quad \quad\quad\quad \quad\quad\quad\quad\quad\quad\quad\quad\quad\quad \mathrm{OR} \quad\nonumber\\
  &&\quad\quad\quad\quad \{\lambda_1,-\lambda_1,\cdots, -\lambda_k,\lambda_k, \cdots,\lambda_j, -\lambda_j,\cdots, \lambda_{N/2},-\lambda_{N/2}\}. \nonumber
 \end{eqnarray}
  Each of the possible target configurations can be identified with a set of $\frac{N}{2}$ numbers that are collected in a vector ${\boldsymbol \sigma}$ whose $j$-th element indicates if the j-th pair $(\lambda_j,-\lambda_j)$ in the target configuration is reversed $(-\lambda_j,\lambda_j)$ ($\sigma_j=1$) or not ($\sigma_j=0$). Each ${\boldsymbol \sigma}$ corresponds to $\left(\frac{N}{2}\right)!$  configurations of rapidities;  instead of summing on $P$ we can sum on ${\boldsymbol \sigma}$ and insert  a factor of $\left(\frac{N}{2}\right)!$. 
 
 Let us now consider the case with $k\neq 0$ given by (\ref{afterLaplaceTlr}). In order for the pole in $s=0$ to be of maximal order, i.e. $\frac{N}{2}+1$,  $\boldsymbol k$ should not only be parity invariant, but also parity invariant within each pair  $(\lambda_j, -\lambda_j)$. Each of these configurations is thus indicated by a vector with $\frac{N}{2}$ elements, ${\boldsymbol \zeta}$, whose $j$-th element indicates if associated to the  $j$-th pair of $\lambda$  $(\lambda_j,-\lambda_j)$ one should sum $(-k,k)$  ($\zeta_j=1$) or $(k,-k)$ ($\zeta_j=0$). 
Then we get
\begin{eqnarray}
&&\sum_{P,Q}\left. \mathrm{Res}\left(\frac{e^{sL}}{s}\prod_{j=1}^N\frac{1}{s-i\sum_{l=j}^N\lambda_{P_l}^Q}\right)\right|_{s=0}=\left(\frac{N}{2}\right)!\sum_{\boldsymbol \sigma, \boldsymbol \zeta}  \mathrm{Res}\left(\frac{e^{sL}}{s^{N/2+1}}\prod_{j=1}^N\frac{1}{s+i((-1)^{\sigma_j} \lambda _j +(-1)^{\zeta _j} k)}\right)\nonumber\\
&&= L^{N/2}\sum_{\boldsymbol \sigma, \boldsymbol \zeta}\prod_{j=1}^{N/2}\frac{1}{i\left( (-1)^{\sigma _j}\lambda _j+(-1)^{\zeta _j}k\right)}\left(1+{\cal O}\left(\frac{1}{L}\right)\right).
\end{eqnarray}
Taking into account that 
 \begin{eqnarray}
 \prod_{j>l=1}^N \left(1-\frac{ic}{\lambda_{P_j}-\lambda_{P_l}}\right)=\prod_{j>l=1}^{N/2} \frac{(\lambda_{j}-\lambda_{l})^2+c^2}{(\lambda_{j}-\lambda_{l})^2}  \frac{(\lambda_{j}+\lambda_{l})^2+c^2}{(\lambda_{j}+\lambda_{l})^2} \left(1+(-1)^{\sigma_l}\frac{ic}{2 \lambda_l}\right)
 \end{eqnarray}
 and putting everything together we find that (\ref{afterLaplaceTlr}), in the zero density limit, is given by
 \begin{eqnarray}
 && I^\leftrightarrows(L)=  \frac{(-2L/c)^{N/2}}{\sqrt{N!}}\frac{1}{\prod_{j=1}^{N/2}\left(\frac{\lambda _j}{c}-\frac{\alpha^2 c}{\lambda_j}\right)\sqrt{1/4+\lambda _j^2/c^2}} \left(1+{\cal O}\left(\frac{1}{L}\right)\right),
\end{eqnarray}
where we recall that $\alpha=\frac{k}{c}$.
According to (\ref{exp_overlap}) the overlap in its continuum form  can  be written as
 \begin{equation}\label{formulakino}
 \langle  0^\leftrightarrows|{\boldsymbol \lambda}\rangle _n= \exp\left\{-\frac{Ln}{2}\left(1+\ln\frac{c}{n}\right)- \frac{L}{2}  \int_0^{\infty} d \lambda \ \rho(\lambda) \ln \left[  \left(  \frac{\lambda }{c}-\alpha^2\frac{c}{\lambda}\right)^2  \left(\frac{1}{4}+\frac{\lambda^2}{c^2}\right) \right]\right\}.
 \end{equation}
 
Let us now compute the  overlap for  $N, L$ finite ($N=2, 4$). For  $N=2$  and a Bethe state of the form $|\lambda_1, -\lambda_1\rangle= |\beta, -\beta\rangle$, with $\beta$ generic, (\ref{afterLaplaceTlr}) becomes
\begin{eqnarray}\label{kinoprimoN2}
&&I^\leftrightarrows(L)=  \frac{2 \beta}{\sqrt{2(4 \beta^2+ c^2)}}\left[\left(1+\frac{ic}{2 \beta}\right)(s^*(\beta)+s(-\beta))+\left(1-\frac{ic}{2 \beta}\right)(s(\beta)+s^*(-\beta))\right],
\end{eqnarray}
where $s(\beta)= \frac{1-e^{-iL(k-\beta)}-iL(k-\beta)}{(k-\beta)^2}$. Enforcing the Bethe equations we get
\begin{eqnarray}\label{kinoN2}
&&I^\leftrightarrows(L)=  \frac{2\sqrt{2}\beta c L}{(k^2-\beta ^2)\sqrt{4 \beta^2 +c^2}}\left[1+\frac{4\sin \frac{L k}{2}}{c L (\beta^2 -k^2)}\left(\sin \frac{L k}{2} (\beta ^2 +k^2)-c k \cos \frac{L k}{2}  \right) \right].
\end{eqnarray} 
As $L\to \infty$, it reduces to
\begin{equation}
I^\leftrightarrows(L)=\frac{\sqrt{2}\beta  L}{(k^2-\beta ^2)\sqrt{1/4+\beta ^2/c^2}}\left(1+{\cal O}(1/L)\right),
\end{equation}
 which is (\ref{formulakino}) for $N=2$.
 
For $N=4$, with a state of the form $|\lambda_1, -\lambda_1,\lambda_2, -\lambda_2 \rangle =|\beta, -\beta, \gamma, -\gamma\rangle $, the leading in $L$ term of the overlap turns out to be of the form
 \begin{equation}
I^\leftrightarrows(L)=\frac{  \sqrt{\frac{2}{3}}   \gamma \beta L^2}{(k^2-\gamma ^2) (k^2-\beta ^2) \sqrt{(1/4+ \gamma^2 / c^2)(1/4+ \beta ^2 / c^2)}}\left(1+{\cal O}(1/L)\right),
\end{equation}
 which is (\ref{formulakino}) for $N=4$.  Since the leading term in $L$ of the overlap, in the finite density cases we have examined,  coincides with its zero density limit, we can safely state that the extensive part of the overlap in the thermodynamic limit is given by   (\ref{formulakino}).

\section{Saddle point distribution  for  a state with oppositely moving BECs in the large quench limit }\label{AppC}

In this section we derive  expression (\ref{pert_res}) for the saddle point  solution $\tilde \rho_s^{(1)}(x)$  in the case of  oppositely moving BECs  in the large quench limit.

Let us consider the non linear integral equation (\ref{sprot2}) as  $\tau\to 0$. For any fixed $x$, with $x>\tau$, the convolution integral gives a subdominant contribution. It is easy to check that  the lowest order in $\tau$ for $a(x, \tau, \alpha)$ is 
\begin{equation}\label{zeroKINO}
a^{(0)}(x, \tau, \alpha)=\frac{\tau^2 x^2}{(x^2-\alpha^2)^2(x^2+\frac{1}{4})};
\end{equation}
in fact the convolution integral, computed with (\ref{zeroKINO}), gives subleading contributions, i.e 
\begin{equation}
\int_{-\infty}^{\infty}\frac{d y}{2 \pi} {\cal K}(y-x) \ln (1+a^{(0)}(y,\tau, \alpha))=\int_{-\infty}^{\infty}\frac{d y}{2 \pi}\ln \left(\frac{\prod_{i=1}^3(y^2+t_i)}{(y^2+1/4)(y-\alpha)^2 (y+\alpha)^2}\right)=0,
\end{equation}
with the roots $t_i$, at the zero-th order in $\tau$, being
\begin{eqnarray}
&& \sqrt{t_1}=\sqrt{t_2}=i \alpha+ O(\tau), \quad \sqrt{t_3}=1/2+ O(\tau).
\end{eqnarray}
The solution to the next order in $\tau$, namely $a^{(1)}(x,\tau, \alpha)$, can be derived inserting the zero-th order term  (\ref{zeroKINO}) in the convolution integral of (\ref{sprot2}) and solving accordingly for $a^{(1)}(x,\tau, \alpha)$. We get
\begin{eqnarray}\label{formfirstorde}
\ln(a^{(1)}(x,\tau, \alpha))&=&\ln(a^{(0)}(x,\tau, \alpha))+\int_{-\infty}^{\infty} \frac{d y}{2 \pi} {\cal K}(y-x) \ln (1+a^{(0)}(y,\tau, \alpha))\nonumber\\
&=&\ln(a^{(0)}(x,\tau, \alpha))+\int_{-\infty}^{\infty} \frac{d y}{2 \pi} {\cal K}(y-x) \ln\left(\frac{\prod_{i=1}^3(y^2+t_i)}{(y^2+1/4)(y-\alpha)^2 (y+\alpha)^2}\right),
\end{eqnarray}
where $t_i$ have  to be kept up to the  first order in $\tau$, i.e. $\sqrt{t_1}=i \alpha +\frac{\tau}{\sqrt{1+4 \alpha^2}}+O(\tau^2)$,   $\sqrt{t_2}=i \alpha -\frac{\tau}{\sqrt{1+4 \alpha^2}}+O(\tau^2)$, $\sqrt{t_3}=1/2+O(\tau^2)$. By making use of the formula 
\begin{equation}
\int_{-\infty}^{\infty} \frac{d y}{2 \pi} {\cal K}(y-x) \ln (y^2+t_1)=\ln(x^2+(1+\sqrt{t_1})^2), \qquad t_1\ \epsilon\  \mathbb{C}, \quad \operatorname{Re}[\sqrt{t_1}]>0,
\end{equation}
(\ref{formfirstorde}) can be rewritten as
\begin{equation}
\frac{a^{(1)}(x,\tau, \alpha)}{a^{(0)}(x,\tau, \alpha)}=\frac{(x^2+(1+\sqrt{t_1})^2)(x^2+(1+\sqrt{t_2})^2)}{(1+(x-\alpha)^2)(1+(x+\alpha)^2)}.
\end{equation}
Substituting (\ref{zeroKINO}), eventually we get
\begin{equation}
a^{(1)}(x,\tau, \alpha)=\frac{\tau^2 x^2}{(x^2-\alpha^2)^2(x^2+\frac{1}{4})}\left(1+\frac{4 \tau (1+x^2+\alpha^2)}{\sqrt{1+4 \alpha^2}  (1+(x-\alpha)^2)(1+(x+\alpha)^2) }\right)
\end{equation}
 and, after applying (\ref{a-rho-rel}), we obtain $\tilde\rho^{(1)}(x,\tau, \alpha)$
\begin{equation}\label{pert_res1}
\tilde\rho^{(1)}(x,\tau, \alpha)=\frac{x^2 \tau ^2\left[6 \tau (1+ x^2+\alpha^2)+\sqrt{1+4 \alpha^2} (1+(x-\alpha)^2)(1+(x+\alpha)^2)\right]}{2 \pi \left[\sqrt{1+4 \alpha^2} (1+(x-\alpha)^2)(1+(x+\alpha)^2) \left(x^2\tau^2 +(x^2-\alpha^2)^2(x^2+\frac{1}{4})\right)+4 x^2\tau^3(1+x^2+\alpha ^2) \right]}\nonumber,
\end{equation}
which  is our perturbative result for $\tilde\rho(x,\tau, \alpha)$ up to $\tau^3$, valid for small $\tau$ and any value of $\alpha$.  

\bibliographystyle{jhep}
\bibliography{leda}

\end{document}